\documentclass[aps,pre,reprint, amsmath, amssymb,superscriptaddress]{revtex4-2}
\usepackage{bm}

\usepackage{graphicx,tikz,placeins}
\usepackage{mathrsfs}
\usepackage[english]{babel}
\usepackage{amsmath,amssymb,amsfonts}
\usepackage{color}
\usepackage[colorlinks=true,linkcolor=blue,urlcolor=blue,citecolor=blue,pdfusetitle]{hyperref}

\DeclareMathOperator{\sech}{sech}

\begin{document}

\title{Characterization and optimization of heat engines: Pareto-optimal fronts and universal features}
%\title{Power, dissipation and fluctuation relations in nonequilibrium heat engines}

\author{Gustavo A. L. For\~ao}
\affiliation{Universidade de São Paulo,
Instituto de Física,
Rua do Matão, 1371, 05508-090
São Paulo, SP, Brazil}

\author{Jonas Berx}
\email[]{jonas.berx@nbi.ku.dk}
\affiliation{Niels Bohr International Academy, Niels Bohr Institute,
University of Copenhagen, Blegdamsvej 17, 2100 Copenhagen, Denmark}

\author{Carlos E. Fiore}
\affiliation{Universidade de São Paulo,
Instituto de Física,
Rua do Matão, 1371, 05508-090
São Paulo, SP, Brazil}

\date{\today}

\begin{abstract}
Characterizing and optimizing nanoscopic heat engines require an appropriate understanding of the interplay between power, efficiency, entropy production and fluctuations. Despite significant recent advancements, including linear stochastic thermodynamics and thermodynamic uncertainty relations, a complete scenario remains elusive.  In this work, we give a further step by showing that, under certain common and general conditions, the heat engine regime can be characterized by the minima of power fluctuations and entropy production, which together delimit its optimal performance—achieved when these conditions are fully satisfied. Conversely, when these conditions are not strictly met, the occurrence of the minimum still approximately describes the system, suggesting a broader range of applicability. Contrasting with most of studies in which the system optimization is carried out solely taking into account the power or efficiency, we introduce a multi-objective optimization framework based on Pareto fronts, also considering the role of fluctuation and dissipation. Our results reveal a general trend: while simultaneous optimization over a few parameters typically yields convex Pareto fronts, these fronts become concave as more parameters are varied freely and non-conservative driving becomes significant. Illustrating our findings, we consider simple two and three state systems as well as richer collective systems, exhibiting novel aspects of optimizations and protocol phase transitions.

\end{abstract}

\maketitle

\section{Introduction}

The study of energy conversion at microscopic scales is a central topic in nonequilibrium thermodynamics~\cite{seifert2012stochastic,landientropy}, the relevance of which encompasses a wide range of systems in physics~\cite{browniancarnot,ciliberto,ciliberto2017experiments}, chemistry~\cite{berton2020thermodynamics}, biology~\cite{kinesin,kinesin2,pedrodelosrios} and quantum technologies~\cite{landientropy}, showcasing the growing interest in understanding the mechanisms behind nonequlibrium thermal machines operating at the nanoscale. Taking into account the unavoidable role of fluctuations and
dissipation in such systems, it is desirable to obtain a comprehensive framework including such effects, as well as their influences on the system performance (e.g. power and/or efficiency). Although remarkable recent
progress has been achieved through, e.g., linear stochastic thermodynamics~\cite{van2005thermodynamic,karel2016prl}, thermodynamic uncertainty relations (TURs)~\cite{barato2015thermodynamic, gingrich, dieball,turbreak} and experimental efforts~\cite{maggi,pal}, a complete scenario remains incomplete.

The purpose of the present contribution is twofold. First, we introduce an alternative approach for characterizing a broad class of nonequilibrium steady-state heat engines. We show that under certain common but general conditions the engine regime can be characterized via the minima of power fluctuations and dissipation (entropy production), where the optimal trade-off between them ensures the global maximization of the power. The generality of our results, exemplified in simpler and more revealing cases such as two-state, three-state and collectively operating systems ~\cite{gatien,filho2023powerful,splitting2024}, 
sheds light on 
an alternative way of portraying heat engines as well as the role of its fluctuations and dissipation.
Aimed at scrutinizing the optimization of heat engines beyond individual quantities, e.g., power~\cite{van2005thermodynamic,tu2008efficiency,seifert2011efficiency,verley2014unlikely,cleuren2015universality,golubeva2012efficiency,esposito2009universality,schmiedl2007efficiency,bonanca2019,campisi2016power} and efficiency~\cite{proesmans2015onsager,ciliberto2017experiments,mamede2021obtaining,filho2023powerful,noa2020thermodynamics,noa2021efficient,splitting2024}, we then consider different kinds of simultaneous optimizations of power, efficiency, dissipation and power fluctuations.
We approach such multi-objective optimization problems via Pareto-optimal fronts~\cite{pareto,berx2024universal,Berx2024}, which provide a more complete picture of engine performance, allowing for the design of heat engines that balance all of the aforementioned quantities in a controlled way.
Results pertaining to entirely different kinds of heat engines reveal general discoveries. While simultaneous optimizations with respect to a few parameters are generally characterized by Pareto fronts assuming a convex shape, they become concave as more parameters are freely varied and asymmetry (here represented by non-conservative drivings) is significant. Such convex and concave fronts are closely related to protocol phase transitions, with power changing smoothly in the former and abruptly in the latter as it increases from its root to its maximum value.

This paper is structured as follows: In Sec.~\ref{2} we introduce the theoretical framework and define the main thermodynamic quantities, whereas Sec.~\ref{theo} presents the relationship between the engine regime and the minimum of power fluctuations and dissipation, as well as a description of our optimization scheme and the analogy to classical thermodynamics. Applications of our framework are shown in Sec.~\ref{case} and  in Sec.~\ref{conclusions}, we draw our conclusions and comment briefly on perspectives.

%The study of energy conversion at microscopic scales is a central topic in nonequilibrium %thermodynamics. Understanding how power, entropy production, and fluctuations interact in small-%scale heat engines is crucial for designing efficient thermal machines, specially after the %discovery \cite{barato2015thermodynamic,gingrich,dieball} and experimental proof \cite{maggi,pal} %of correlations and (almost \cite{turbreak})  universal constraints on these quantities, the so %called Thermodynamical Uncertainty Relation (TUR). 

%While previous works have explored performance optimization through power-efficiency trade-offs and %fluctuation theorems, our work establishes a new important constraint: under certain (common) %conditions, the heat engine regime is bounded between the minimum of power fluctuations and the %minimum of entropy production.

%... Minimums in entropy production were already seen in linear stochastic thermodynamics. Here, we %present an extension to master equation ruled processes.

\section{General description and Thermodynamics}\label{2}
We consider a discrete Markovian system placed in contact with heat and particle reservoirs at inverse temperatures $\beta_\nu$ and chemical potentials $\mu^{(\nu)}$, with $\nu = 1(2)$ corresponding to the cold~(hot) reservoirs. The thermodynamic affinity associated to the transition from state $j$ to state $i$ due to the reservoir $\nu$ is given by $d_{ij}^{(\nu)} = \Delta\epsilon_{ij}^{(\nu)} - \mu^{(\nu)}\,{\bar n}_{ij}+F^{(\nu)}_{ij}$, where $\Delta\epsilon^{(\nu)}_{ij}=\epsilon^{(\nu)}_{i}-\epsilon^{(\nu)}_{j}$ is the energy difference between above states,  ${\bar n}_{ij}$ and $F^{(\nu)}_{ij}$ denote the number of particles exchanged and the influence of an external  force, respectively. Transition rates are then expressed in Kramers' form, $W_{ij}^{(\nu)}=\Gamma  e^{-\frac{\beta_\nu}{2}d_{ij}^{(\nu)}}$. For simplicity, we will set $\Gamma=1$. The time evolution of the probability is governed by the master equation
\begin{equation}
\dot{p}_i = \sum_\nu\sum_{j\neq i}\{W_{ij}^{(\nu)}\,p_j - W_{ji}^{(\nu)}\,p_i\}\,,
\label{me}
\end{equation}
where the term on the right side corresponds to the probability current  $J_{ij}^{(\nu)} = W_{ij}^{(\nu)}\,p_j - W_{ji}^{(\nu)}\,p_i$. We are interested
in the nonequilibrium steady state (NESS) properties, characterized by the set of probabilities  $\{p^{\rm st}_l\}$ in which $\sum_{\nu}\sum_{j\neq i} \{W_{ij}^{(\nu)}\,p^{\rm st}_j - W_{ji}^{(\nu)}\,p^{\rm st}_i\}\ =0$.

The model thermodynamics  is set up as follows. The steady-state entropy production $\langle \dot{\sigma} \rangle$ \cite{schnakenberg,vandenbroeck15}, 
 the power $\langle {\cal P}\rangle$ and power fluctuations $\textrm{var}({\cal P})$, evaluated in the NESS, are defined as follows:
\begin{eqnarray}
 \langle \dot{\sigma} \rangle= \sum_{\nu}\sum_{i<j}J_{ij}^{(\nu)}\ln\frac{W^{(\nu)}_{ij}}{W^{(\nu)}_{ji}}=-\sum_{\nu}\beta_\nu\langle \dot{{Q}}_\nu\rangle,
 \label{theoriticalep}
\end{eqnarray}
where $\langle \dot{Q}_\nu \rangle=\sum_{i<j}d_{ij}^{(\nu)}J_{ij}^{(\nu)}$ denotes the exchanged heat with the $\nu$-th thermal bath, expressed in terms of affinity $d_{ij}^{(\nu)}$ and $J_{ij}^{(\nu)}$. From the first law of thermodynamics, the expression for $\langle {\cal P}\rangle $
is obtained 
\begin{equation}
\label{power}
    \langle {\cal P}\rangle = -\left(\langle \dot{Q}_1 \rangle + \langle \dot{Q}_2 \rangle\right).
\end{equation}
The evaluation of power fluctuations $  \textrm{var}({\cal P})$ is more involved and for this reason will be considered via two different (but equivalent) approaches. More specifically, we write $\textrm{var}({\cal P}) =\sum_{(\nu,\nu')}\left[\langle \dot{Q}_\nu\dot{Q}_{\nu'}\rangle-\langle \dot{Q}_\nu\rangle\langle\dot{Q}_{\nu'}\rangle\right]$, where each term in the square brackets denotes the covariance. In  order to obtain
$\langle \dot{Q}_\nu\dot{Q}_{\nu'}\rangle-\langle \dot{Q}_\nu\rangle\langle\dot{Q}_{\nu'}\rangle$,  we extend the framework proposed in Ref.~\cite{busiello2022hyperaccurate} for the two reservoirs context. For that, we consider a stochastic trajectory of length $M$ performed by a discrete-state system in the discrete time interval $t \in [0,1,2,...,t_f]$, characterized by the set of visited states, $\{ x_i \}_{i = 0,\dots,M}$,  the expression for the heat current is given by
\begin{equation}
    {\dot Q}_{\nu} = \frac{1}{t_f} \sum_{ij} d^{(\nu)}_{ij} n^{(\nu)}_{ij},
    \label{current}
\end{equation}
where $n^{(\nu)}_{ij} = \sum_{k=0}^{M-1} \delta_{x_k,i} \delta_{x_{k+1},j}$ is the number of jumps from the state $j$ to $i$ up to time $t_f$ in contact with thermal bath $\nu$. From Eq.~(\ref{current}) and by averaging over all trajectories, one finds that
\begin{equation}
    \langle {\dot Q}_\nu \rangle  = \frac{1}{t_f} \sum_{i<j} d^{(\nu)}_{ij}J^{(\nu)}_{ij},
    \label{av1}
\end{equation}
and
\begin{equation}
\langle \dot{Q}_\nu\dot{Q}_{\nu'}\rangle-\langle \dot{Q}_\nu\rangle\langle\dot{Q}_{\nu'}\rangle     = \frac{1}{t_f^2} \sum_{iji'j'} d^{(\nu)}_{ij} d^{(\nu')}_{i'j'} C^{(\nu,\nu')}_{iji'j'},
\label{sigma1}
\end{equation}
respectively, where $J^{(\nu)}_{ij}=\langle n^{(\nu)}_{ij}\rangle - \langle n^{(\nu)}_{ji} \rangle$ and
$C^{(\nu,\nu')}_{iji'j'} = \langle n^{(\nu)}_{ij} n^{(\nu')}_{i'j'} \rangle - \langle n^{(\nu)}_{ij} \rangle \langle n^{(\nu')}_{i'j'} \rangle$, and where the
anti-symmetric property $d^{(\nu)}_{ij}=-d^{(\nu)}_{ji}$
was taken into account in Eq.~\eqref{av1}. By considering it
 in Eq.~\eqref{sigma1} as well, we can restrict the summation over all indices $i < j$ and $i' < j'$, in such a way that
 \begin{equation}
\langle \dot{Q}_\nu\dot{Q}_{\nu'}\rangle-\langle \dot{Q}_\nu\rangle\langle\dot{Q}_{\nu'}\rangle= \sum\limits_{\substack{i<j\\i'<j'}} d^{(\nu)}_{ij} d^{(\nu')}_{i'j'} \mathcal{M}^{({\nu},{\nu'})}_{iji'j'},
\label{var2}
\end{equation}
where
$\mathcal{M}^{({\nu},{\nu'})}_{iji'j'}= C^{({\nu},{\nu'})}_{iji'j'} + C^{({\nu},{\nu'})}_{iji'j'} - C^{({\nu},{\nu'})}_{jii'j'} - C^{({\nu},{\nu'})}_{ijj'i'}$.

\textcolor{black}{In Appendix~\ref{covv}, we describe how to compute $J_{ij}^{(\nu)}$ and $C^{({\nu},{\nu'})}_{iji'j'}$ from
the transition rates. Alternatively, the  evaluation of $\textrm{var}({\cal P})$ via large-deviation method, by following the ideas of Refs.~\cite{touchette2009large, kumar2011thermodynamics}, in which the scaled cumulant generating function is determined by the largest eigenvalue \( \lambda_p(\alpha) \) of \( M_p(\alpha) \). From \( \lambda_p(\alpha) \), the variance is then given by
\begin{equation}
    \textrm{var}({\cal P}) = \left. \frac{\partial^2 \lambda_p(\alpha)}{\partial \alpha^2} \right|_{\alpha=0}.
\end{equation}
More details about the method can also be found in Appendix \ref{covv}.}
\section{Characterization and optimization of heat engines}\label{theo}
Throughout the rest of this work we will be concerned with two main issues:
the characterization of heat engines via the mutual relations between power, dissipation and power fluctuations, and reliable strategies for the simultaneous optimization of these quantities.

\subsection{Power, fluctuations and dissipation}
\label{pfd}
A heat engine typically  converts a partial amount of heat extracted from the hot reservoir  $\langle \dot{Q}_2 \rangle>0$ into power output $\langle\mathcal{P}\rangle<0$, where we choose positive quantities to flow \emph{into} the system. It differs from a heat pump in which $\langle\mathcal{P}\rangle>0$ is partially converted into heat (or particles) from the cold reservoir to the hot one. In both cases, performance
can be quantified by the efficiency, defined as $\eta = -\langle{\cal P}\rangle/\langle \dot{Q}_2 \rangle$, ranging from $0 \leq \eta \leq \eta_c$ (heat engine) and  $\eta_c< \eta \leq \infty$ (heat pump), where $\eta_c = 1-\beta_2/\beta_1$ is the Carnot efficiency. The regime where $\eta < 0$ denotes a ``dud", in which no useful power is produced. Throughout this work, we will focus solely on the heat engine regime.

This section is aimed at showing that \textcolor{black}{in the absence of driving (biased) forces, i.e., we set here $F^{(\nu)}_{ij}=0$}, this regime is constrained between the minimum of power fluctuations and the dissipation (characterized by the entropy production) for a relatively broad class of systems. Consider a reversible Markovian system in its stationary state, interacting with two thermal baths at inverse temperatures $\beta_1$ and $\beta_2$. Let $d_{ij}^{(\nu)}$ denote the thermodynamic affinity associated with the transition from state $j$ to state $i$ due to the thermal reservoir $\nu$. If there exist configurations such that $d_{ij}^{(1)} = d_{ij}^{(2)}$ and $\beta_1\,d_{ij}^{(1)} = \beta_2\,d_{ij}^{(2)}$ for all $i$ and $j \neq i$, then there are two solutions for the condition $\langle{\cal P}\rangle=0$, each one corresponding to the global minimum of either the power fluctuations or the entropy production. To establish this result, we consider a class of Markovian systems in which the condition $(W_{ij}^{(1)} + W_{ij}^{(2)})\,p^{\rm st}_j = (W_{ji}^{(1)} + W_{ji}^{(2)})\,p^{\rm st}_i$ holds in the NESS regime. This implies that the currents satisfy
\begin{equation}
    J_{ji}^{(1)} = J_{ij}^{(2)} = -J_{ji}^{(2)}.
    \label{eq:currents}
\end{equation}
\newline
Note that this result  represents a form of stalled current \cite{espositopolettini}, as \( J_{ij}^{(1)} + J_{ij}^{(2)} = 0 \), and can be viewed as an extension of the concept of reversibility to far-from-equilibrium systems. 
As a consequence, thermodynamic quantities are then fully determined by the probability current from \emph{a single bath alone}. Plugging equation~\eqref{eq:currents} into~\eqref{theoriticalep} and~\eqref{power}, one directly arrives at the expressions 
\begin{equation}
    \langle \dot{\sigma} \rangle = \sum_{i<j}\,(\beta_2\,d_{ij}^{(2)} - \beta_1\,d_{ij}^{(1)})\,J_{ij}^{(1)}\,,
    \label{ep}
\end{equation}
and
\begin{equation}
    \langle \mathcal{P} \rangle =\sum_{i<j} \,(d_{ij}^{(2)} - d_{ij}^{(1)})\,J_{ij}^{(1)}.
    \label{pow}
\end{equation}
The computation of the power fluctuations, i.e., the power variance is more involved. As a consequence of equation~\eqref{eq:currents}, it follows that $\mathcal{M}^{(2,2)}_{iji'j'}=\mathcal{M}^{(1,1)}_{iji'j'}$ and $\mathcal{M}^{(1,2)}_{iji'j'}=\mathcal{M}^{(2,1)}_{iji'j'}=\mathcal{M}^{(1,1)}_{iji'j'}$, and thus
\begin{widetext}
\begin{equation}
    \textrm{var}\,(\mathcal{P}) =\sum_{\substack{i<j\\i'<j'}}\left(d_{ij}^{(2)}d_{i'j'}^{(2)} + d_{ij}^{(1)}d_{i'j'}^{(1)}-d_{ij}^{(1)}d_{i'j'}^{(2)}-d_{i'j'}^{(1)}d_{ij}^{(2)}\right)\mathcal{M}^{(1,1)}_{ijij}\,.
    \label{var}
\end{equation}
\end{widetext}
Clearly, $\textrm{var}\,(\mathcal{P}) = 0$ and $\langle \mathcal{P} \rangle = 0$ when $d_{ij}^{(2)} = d_{ij}^{(1)}$ and $d_{i'j'}^{(2)} = d_{i'j'}^{(1)}$, with 
$\langle \dot{\sigma} \rangle = \sum_{i<j}\,(\beta_2\, - \beta_1\,)d_{ij}^{(1)}\,J_{ij}^{(1)}\ge 0$, proving the first part of the statement. For the remaining part, we use the fact that the system is completely defined by its temperature and affinities. As such, when the system satisfies the condition $\beta_1\,d_{ij}^{(1)} = \beta_2\,d_{ij}^{(2)}$ for all $i$ and $j \neq i$, it follows that $J_{ij}^{(1)} = J_{ij}^{(2)}$. Combining this with equation~(\ref{eq:currents}), we conclude that the only consistent solution is $J_{ij}^{(1)} = J_{ij}^{(2)} = 0$ for all $i \neq j$. In this case, all probability currents vanish, and consequently $\langle \dot{\sigma} \rangle = 0$ and $\langle \mathcal{P} \rangle = 0$.

To assess the validity of these results, a few brief remarks are in order.  
First, $J_{ij}^{(1)} =- J_{ij}^{(2)} $ is closely related to the condition
 $\Tilde{W}_{ij}\Tilde{W}_{jk}\dots\Tilde{W}_{Ni} =\Tilde{W}_{iN}\Tilde{W}_{kj}\dots\Tilde{W}_{ji}$,, with $\Tilde{W}_{ij}=\sum_\nu W^{(\nu)}_{ij}$, also referred to as the Kolmogorov criterion~\cite{kelly2011reversibility}. As shown in Appendix.~\ref{proof}, for any system coupled to two thermal baths in which $d^{(\nu)}_{ij}=-d^{(\nu)}_{ji}$ \cite{horowitz1}, this criterion holds whenever the condition $\prod_{\gamma}W_{lm}^{(1)} \cdot \prod_{\gamma}W_{lm}^{(2)} = 1$ is satisfied, where $\gamma$ denotes an arbitrary closed path in configuration space. Second, they are always verified for the class
of systems in which the conditions $d_{ij}^{(1)} = d_{ij}^{(2)}$ and $\beta_1 \,d_{ij}^{(1)} = \beta_2\,d_{ij}^{(2)}$ are met (for all $i$ and $j \neq i$), irrespective the number of states.

Third, since it is not a trivial task to analytically verify the above conditions in systems with many states, in the next section we examine different models (two- and three-state systems) where these conditions are either fully satisfied or only partially met. Even in cases where they are not strictly fulfilled (e.g., when \( F_{ij}^{(\nu)} \ne 0 \)), we find that the power zeroes and the minima of fluctuations and entropy production can still occur close to each other, suggesting that this way of characterizing the engine regime has a broader range of applicability.

\subsection{Optimization and Pareto-optimal trade-offs}

One of the remarkable applications of nonequilibrium thermodynamics is to explore reliable strategies for optimizing nanoscopic heat engine performances. Most studies typically deal with individual optimizations of, e.g., power or efficiency~\cite{gatien,filho2023powerful,mamede2023,karel2016prl}. Nevertheless, $\langle \cal P\rangle, \, \eta,$ as well as $\langle \dot{\sigma}\rangle$ and $\textrm{var}({\cal P})$ are not independent but are intrinsically related through the underlying system configuration~\cite{PhysRevLett.120.190602} in such a way that individual optimizations in general are not the best strategy for improving system performance. Although leading to remarkable improvements, simultaneous optimization is not a trivial task. 

Pareto optimization offers a powerful framework for exploring trade-offs between interdependent quantities, giving rise to Pareto fronts—hypersurfaces that represent the set of optimal solutions in a multi-objective optimization problem. It deals with competing objectives where improving one objective necessarily comes at the expense of others. In this work, we will consider four-dimensional multi-function Pareto optimization where the dissipation $\langle \dot{\sigma}\rangle$ and power fluctuations ${\rm var}({\cal P})$ are minimized while the absolute power $\langle \cal P\rangle$ and efficiency $\eta$ are maximized.

{\color{black}Moving along such a Pareto hypersurface amounts to changing the state, i.e., all of the underlying system parameters of our engine in an optimal manner and leads to different but equally optimal engine configurations. The protocol by which this change is made can be described by a set of control parameters $\Lambda = \{\lambda_k\geq 0:\, k=1\dots,K\}$, with $K$ the number of objectives and where $\sum_k \lambda_k =1$ without loss of generality~\cite{sole}. During the protocol, a free energy-like functional $\Omega = \sum_k \lambda_k X_k$ is minimized, with $X_k$ the optimization objectives (power, efficiency, fluctuations, etc.). Geometrically, the weights $\lambda_k$ fix the slope of a tangent hyperplane to the front, and as $\Lambda$ is varied this hyperplane is ``rolled'' along the convex Pareto hull, thereby tracing out every point on the front in exactly the same way that a Legendre–Fenchel transform maps between energy and entropy. 

For simplicity, let us take $\Omega = \lambda X_1 + (1-\lambda) X_2$, $\lambda\in[0,1]$. When $\lambda=0$, the optimal design minimizes $X_2$; when $\lambda=1$ it minimizes $X_2$. As $\lambda$ increases, the optimal solution moves smoothly from the $X_2$-extremal toward the $X_1$-extremal solution along the Pareto front. Treating one objective--say $X_1$-- as an order parameter then describes the protocol taking the solution from point A to point B along the Pareto front. Sudden jumps or kinks in $X_1(\lambda)$ mirror phase transitions in classical thermodynamics, where a system jumps between two coexisting phases at a critical conjugate variable (e.g. temperature or chemical potential). 

By means of example, let us set $X_1 = U$ (energy) and $X_2 = -S$ (entropy) in a classical system, such that we minimize the energy and maximize the entropy; we recover the Helmholtz free energy in the canonical ensemble, $F = \Omega/\lambda = U - T S$, with temperature $T = (1-\lambda)/\lambda$ playing the role of a control parameter. Non-convex “intrusions” in the microcanonical entropy-energy curve—unstable regions of negative heat capacity—are replaced by a convex hull in the canonical ensemble, giving the familiar latent heat jump at the liquid–gas transition. Likewise, any local concavity in a Pareto front signals a discontinuous (first-order) transition in engine design: at a critical $\lambda_c$ two distinct designs become equally optimal under $\Omega$, and the system “jumps” from one to the other. In contrast, a globally convex Pareto front yields smooth, continuous changes in the order parameters—analogous to second-order phase transitions (when there are kinks in the front) or the absence of any transition altogether. Thus the convexity properties of Pareto fronts directly dictate whether an engine’s optimal design varies smoothly with the chosen trade-off parameters or whether it undergoes sharp, phase-transition–like reorganizations. 

In the context of heat engines, Pareto trade-offs between, e.g., power and efficiency exhibiting local non-convex behavior entail a sharp transition in the optimal engine design; one can view $\lambda$ as the optimal protocol in design space, interpolating between designs with, e.g., maximal power or maximal efficiency. Similar to the aforementioned classical example where the control parameter (temperature), was proportional to the rate of change of the energy with the entropy, the control parameter for the power-efficiency trade-off is related to the rate by which the power changes when tuning the efficiency. Including additional optimization objectives, e.g., $\textrm{var}(\mathcal{P})$ is then straightforward: one introduces control parameters $\lambda_1$ and $\lambda_2$, which then represent respectively the rate by which the power changes when tuning either the efficiency or the fluctuations, and Pareto optimization is then similar to minimizing the Landau free energy in grand canonical classical systems.}

In addition to a full four-dimensional optimization, we shall generally consider pairwise trade-offs by projecting the full Pareto-optimal solutions onto a two-dimensional subspace. This subspace is then not necessarily Pareto-optimal anymore with respect to the two chosen objectives, and the required pairwise Pareto fronts can subsequently be obtained through marginalization. Alternatively, the pairwise fronts can be found through a direct pairwise Pareto optimization, without the need for the larger superspace, and we will compare the projective approach with direct pairwise optimization.

{\color{black}Our numerical Pareto optimization is based on the NSGA-II genetic algorithm~\cite{pareto}, using concepts from evolutionary biology to determine optimal parameter combinations that simultaneously optimize multiple objectives while strictly obeying imposed constraints. A brief but more detailed description of the algorithm is given in Appendix~\ref{app:NSGA}.} 
This approach allows for a better global exploration of the solution space while avoiding the need for a reward function with model-dependent hyperparameters, as is common in, e.g., reinforcement learning approaches~\cite{erdman2023pareto}.

Finally, we will also consider some special points of interest on the pairwise Pareto fronts: the \emph{efficiency} at maximal power $\eta_{MP}$, \emph{dissipation} at maximal power (DMP) $\langle \dot{\sigma}\rangle_{MP}$ and power \emph{fluctuations} at maximal power (FMP) $\mathrm{var}(\mathcal{P})_{MP}$. Note that generally the efficiency is rescaled by the Carnot efficiency, $\eta_c$, when considering different trade-offs. When performing the Pareto optimization, one then needs to take special care in formulating the problem, since optimizing with respect to $\eta$ or to $\eta/\eta_c$ leads
to different results in cases where reservoir temperatures are included as optimization parameters. 

\section{Applications}\label{case}

\begin{figure}
    \centering
    \includegraphics[width=0.9\linewidth]{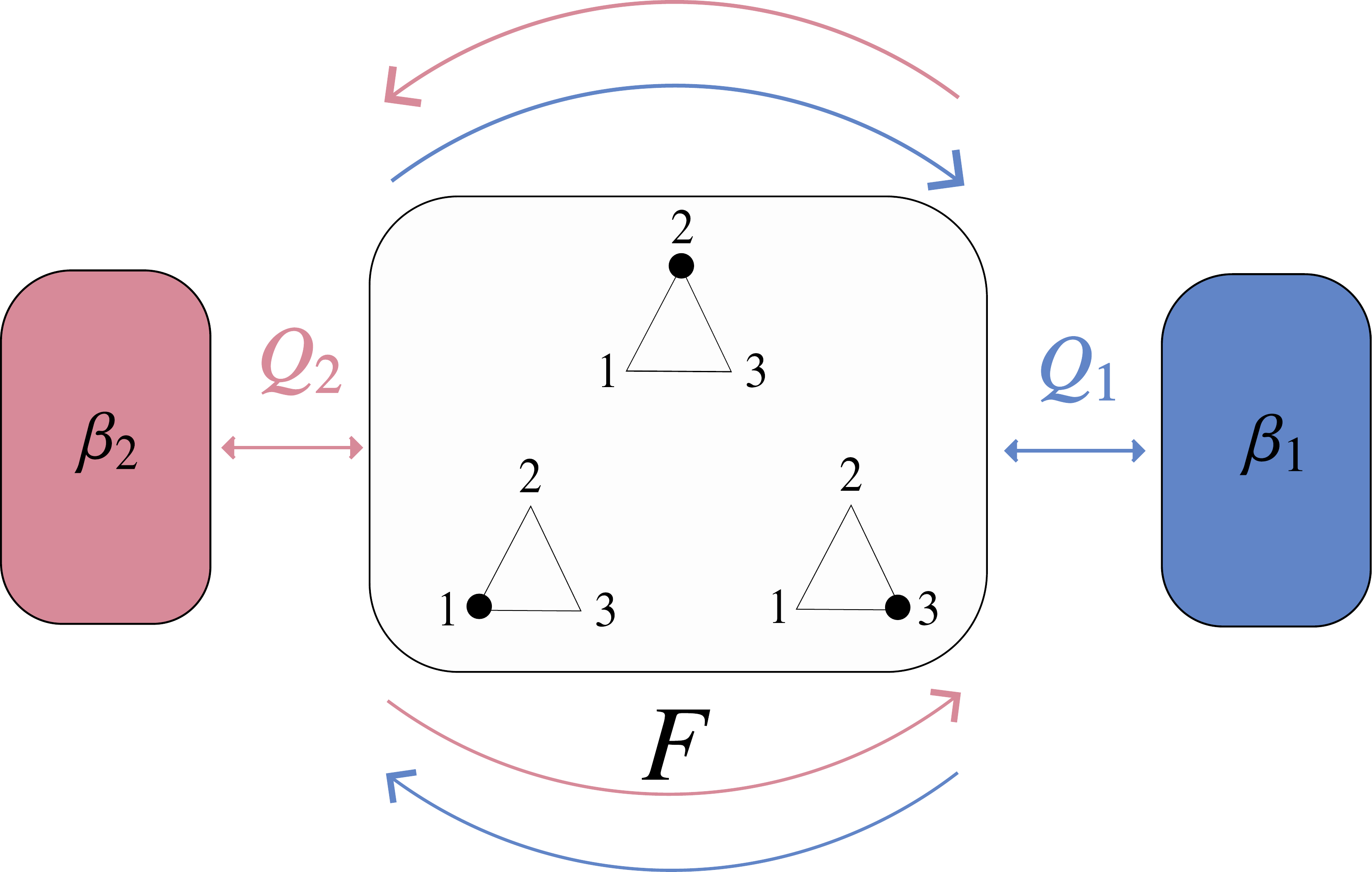}
    \caption{\textcolor{black}{Schematics of the driving operation for the three-state system  in contact with the cold (left) and hot (right) thermal baths. Arrows denote the spin transitions $1\rightarrow 2\rightarrow 3\rightarrow 1$ favored in the clockwise (counterclockwise) by the non-conservative driving of strength $F$ provided through the coupling with the cold~(hot) thermal baths. The protocol is similar for the collective system by replacing the sequence $(1\rightarrow 2\rightarrow3)$ to a local spin $s_i$ in which $(-1\rightarrow 0\rightarrow 1)$.}}
    \label{scheme}
\end{figure}

We consider two classes of systems: those composed of a single unit and those consisting of many interacting units. While the former is simpler and serves as idealized setting for verifying previous results, the latter exhibits novel features attributed solely to the collective behavior~\cite{filho2023powerful,feyisa2024trapped,PhysRevApplied.21.044050,PhysRevA.105.043708}, such as phase transitions without equilibrium
or nonequilibrium analogs~\cite{splitting2024}.
Although similar results can be obtained for $\mu_\nu \neq 0$, our analyses
will be carried out for the simplest case  $\mu_\nu = 0$ (hence  affinities become $d_{ij}^{(\nu)} = \Delta \epsilon_{ij}^{(\nu)}+F^{(\nu)}_{ij}$). \textcolor{black}{Transition rates for all cases are listed in Appendix \ref{rates}.}
\subsection{Non-interacting systems}

\begin{figure*}[tp]
    \includegraphics[width=\linewidth]{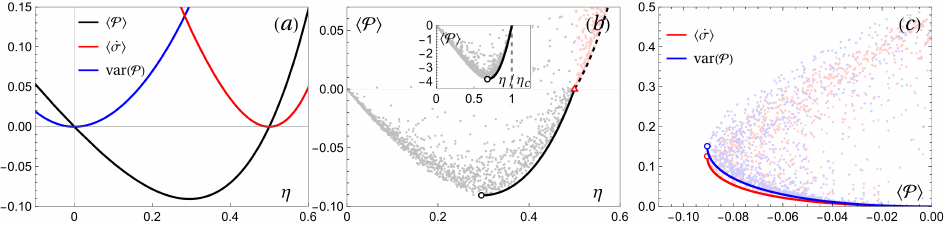}
     \caption{Two-state thermodynamic observables and trade-offs. {\bf(a)} Plot of the power, dissipation and power fluctuations as a function of efficiency $\eta$ for fixed $\beta_1 = 2$, $\beta_2 = 1$ and $\epsilon^{(2)} = -3.5$. {\color{black} Note that the heat engine regime, where $\langle\mathcal{P}\rangle <0$, is exactly bounded by both the minima of power fluctuations and of the dissipation}; {\bf (b)} pairwise power-efficiency Pareto front transitioning from the engine (black full line) to the heat pump (dashed line, red dots) regime at $\eta = \eta_C$ (red open triangle). Dots result from four-dimensional Pareto optimization. \emph{Inset:} Pairwise optimization of $\langle \mathcal{P}\rangle$--$\eta/\eta_c$, only in the engine regime. {\bf (c)} $\langle \mathcal{P}\rangle$--${\langle {\dot \sigma} \rangle}$ (red) and $\langle \mathcal{P}\rangle$-${\rm var}\mathcal{(P)}$ (blue) trade-offs. In panels (b, c), reservoir temperatures $\beta_1=2$, $\beta_2=1$ are fixed while $\epsilon^{(1)},\,\epsilon^{(2)} \in[-7,0]$) for the inset of panel (b), and $\beta_2 \in [0,\beta_1]$ with $\beta_1 = 2$. Open circles indicate the EMP in panel (b), and the DMP, FMP in panel (c).
     }
    \label{pareto2level}
\end{figure*}

\textit{Two-state systems—} As the first (and simplest) application, we consider a generic two-state system, \textcolor{black}{composed of a  single unit which can be in a low energy ($i=1$) or high energy ($i=2$) state, with corresponding energies $0$ and $\epsilon^{(\nu)}$, respectively}. Despite its simplicity, different paradigmatic examples such as paramagnetic systems or quantum dots~\cite{fiorek,harunari2020maximal} can be modeled as two-state systems. Affinities then assume the form $d^{(\nu)}_{21}= \epsilon^{(\nu)}+F^{(\nu)}_{21}$ and $d^{(\nu)}_{12}=- \epsilon^{(\nu)}+F^{(\nu)}_{12}$, respectively. 
As the system presents only a single independent flux, the thermodynamic quantities studied acquire simple forms. For  $F^{(\nu)}_{21}=F^{(\nu)}_{12}=0$, they are given by
$\langle \mathcal{P} \rangle = (\epsilon^{(1)} - \epsilon^{(2)})\,J_{12}^{(1)}$ and $ \langle \dot{\sigma} \rangle = (\beta_1\,\epsilon^{(1)} - \beta_2\,\epsilon^{(2)})\,J_{12}^{(1)}$, where
$J_{12}^{(1)}$ reads
\begin{equation}
J_{12}^{(1)}=\sinh{\left(\frac{\beta_1\epsilon^{(1)}-\beta_2\epsilon^{(2)}}{4}\right)} \sech{\left(\frac{\beta_1\epsilon^{(1)}+\beta_2\epsilon^{(2)}}{4}\right)}.
\label{flux}
\end{equation}
Similarly, the power fluctuations assume the form $\textrm{var}\,(\mathcal{P}) =(\epsilon^{(2)} - \epsilon^{(1)})^2\,\mathcal{M}^{(1,1)}_{1212}$, where $\mathcal{M}^{(1,1)}_{1212}$ is given by
\begin{equation}
    \label{vari}
    \begin{split}
        \mathcal{M}^{(1,1)}_{1212}&=\frac{1}{8}\left(2+\cosh{(\beta_1 \epsilon^{(1)}})+\cosh{(\beta_2 \epsilon^{(2)})}\right)\times\\
        &\sech{\left(\frac{\beta_1\epsilon^{(1)}-\beta_2\epsilon^{(2)}}{4}\right)} \sech^3{\left(\frac{\beta_1\epsilon^{(1)}+\beta_2\epsilon^{(2)}}{4}\right)}\,.
    \end{split}
\end{equation}
Results for the two-state model can be seen in Fig.~\ref{pareto2level}(a). In accordance with Eqs.~\eqref{pow} and~\eqref{var}, the roots of $\langle{\cal P}\rangle$ are given by $\epsilon^{(2)} = \epsilon^{(1)}$ and $\epsilon^{(2)} = \frac{\beta_1}{\beta_2} \epsilon^{(1)}$ and $\langle{\cal P}\rangle<0$ for $\epsilon^{(2)}<\epsilon^{(1)}<\beta_2\epsilon^{(2)}/\beta_1 $. The former and latter value correspond to the minima of $\textrm{var}\,(\mathcal{P})$ and $\langle \dot{\sigma} \rangle$, respectively. Since there is only one independent flux, the efficiency acquires the simple form $\eta=1-\epsilon^{(1)}/\epsilon^{(2)}$, consistent with the heat engine regime being constrained between $\eta=0$ at \(\epsilon^{(1)} = \epsilon^{(2)}\) and the Carnot efficiency \(\eta_c = 1 - (\beta_2 / \beta_1)\) when \(\epsilon^{(1)} = (\beta_2 / \beta_1) \, \epsilon^{(2)}\). 
 
Despite  the individual maximization of efficiency being as simple as approaching the limit \(\epsilon^{(1)} \to (\beta_2 / \beta_1) \, \epsilon^{(2)}\), a more complete understanding emerges by jointly optimizing power, efficiency, fluctuations, and the power dissipation. The resulting Pareto fronts from this full four-dimensional optimization are shown in Fig.~\ref{pareto2level}(b,c) as dots, each representing a solution that is Pareto-optimal in the full multidimensional space. These solutions are then projected onto two-dimensional subspaces defined by $\langle \mathcal{P}\rangle$ and a second observable $X\in \{\eta, \eta/\eta_c,{\langle {\dot \sigma} \rangle}, {\rm var}\mathcal{(P)} \}$. However, once projected, these four-dimensional solutions are generally not Pareto-optimal with respect to the specific pairwise trade-off in the reduced space. Instead, the optimal trade-offs within each two-dimensional subspace are captured by the continuous lines, which represent Pareto fronts obtained either by direct pairwise optimization or by marginalizing over the higher-dimensional solutions, shown by full lines in Fig.~\ref{pareto2level}(b,c). These pairwise Pareto fronts identify the most significant trade-offs and always outperform the projected four-dimensional results in their respective subspaces. 

Importantly, each pairwise front is strictly optimal within the interval $0<X<X_{m}$, where $X_{m}$ denote the respective optimal value for objective $X$ (open symbols $\circ$ in Fig.~\ref{pareto2level}(b,c)). Moreover, across all considered cases—including the non-driven scenario—the Pareto fronts display qualitatively similar, convex shapes. However, the performance of the system varies depending on whether the optimization is performed with fixed parameters—namely, $\{\epsilon^{(2)}, \beta_1, \beta_2\}$, $\{\beta_1, \beta_2\}$, or none at all. In the former case, the $\langle{\cal P}\rangle-\eta $ Pareto fronts follow Eqs.~\eqref{pow}, as shown in Fig.~\ref{pareto2level}(a,b).  
%Improved optimization can be achieved by varying both individual energies (main) and all parameters (inset). 

As a final comment, it is worth pointing out that the region where $\eta<\eta_{MP}$ is also suboptimal because it is bounded by a different pairwise optimization, in which  both $\langle \mathcal{P}\rangle$ and $\eta$ are simultaneously \emph{minimized}, i.e., where the absolute power is maximal but the efficiency is minimized. Since this is not a physically interesting region for applications, we do not comment on it any further but it is worth mentioning that such trade-offs emerge naturally by projecting on a lower-dimensional subspace, and they can be determined more accurately by performing separate pairwise optimizations. Likewise, we are not interested in bounds that maximize dissipation and/or power fluctuations while minimizing $\langle{\cal P}\rangle$, since typically one is interested into suppressing fluctuations and lowering dissipation.

\textcolor{black}{We close this section by extending the above analysis for $F^{(\nu)}_{ij}\neq 0$ and $\Gamma_1\neq \Gamma_2$. The former 
is inspired by previous works  \cite{gatien,filho2023powerful,mamede2023,splitting2024},
in which one includes} a non-conservative driving with $F^{(\nu)}_{ij}$, defined as follows: $F^{(\nu)}_{21}=(-1)^{1+\nu}F$ and  $F^{(\nu)}_{21}=- F^{(\nu)}_{12}$. Expressions are obtained as before and are given by $\langle {\cal P}\rangle=( \epsilon^{(1)} - \epsilon^{(2)}-2F)J_{12}^{(1)}$, 
 $\langle {\dot \sigma}\rangle=\left( \beta_1\epsilon^{(1)} - \beta_2\epsilon^{(2)}-(\beta_1+\beta_2)F\right)J_{12}^{(1)}$ and $\textrm{var}\,(\mathcal{P}) =(\epsilon^{(1)} - \epsilon^{(2)}-2F)^2\,\mathcal{M}^{(1,1)}_{1212}$, where
 $J_{12}^{(1)}$ and $\mathcal{M}^{(1,1)}_{1212}$
read
\begin{equation}
    \begin{split}
        J_{12}^{(1)}&= \sinh{\left(\frac{\beta_1 (\epsilon^{(1)}-F)-\beta_2 (\epsilon^{(2)}+F)}{4} \right)}\times\\
        &\sech{\left(\frac{\beta_1( \epsilon^{(1)}-F)+\beta_2 (\epsilon^{(2)}+F)}{4}\right)}
    \end{split}
\end{equation}
and
\begin{widetext}
\begin{equation}
    \mathcal{M}^{(1,1)}_{1212}=
    \frac{e^{-\frac{1}{2} \beta_1 (F+\epsilon^{(1)})} \left(\left(e^{2 \beta_1\epsilon^{(1)}}+4 e^{\beta_1
   (F+\epsilon^{(1)})}+e^{2 \beta_1 F}\right) e^{\beta_2 (F+\epsilon^{(2)})}+e^{\beta_1\epsilon^{(1)}+2
   \beta_2\epsilon^{(2)}+\beta_1 F+2 \beta_2 F}+e^{\beta_1 (F+\epsilon^{(1)})}\right)}{\left(e^{\frac{1}{2} (\beta_1 
   \epsilon^{(1)}+\beta_2 (F+\epsilon^{(2)}))}+e^{\frac{\beta_1
   F}{2}}\right)^2 \left(2 e^{\frac{1}{2} \beta_2 (F+\epsilon^{(2)})} \cosh \left(\frac{1}{2} \beta_1 (F-\epsilon^{(2)})\right)+e^{\beta_2 (F+\epsilon^{(2)})}+1\right)},
\end{equation}
\end{widetext}
respectively. The efficiency now reads $\eta=1-(\epsilon^{(1)}-F)/(\epsilon^{(2)}+F)$,
consistent with the engine regime delimited by $\epsilon^{(1)}-\epsilon^{(2)}=2F$, where $\eta=0$ and $\beta_1(\epsilon^{(1)}-F)=\beta_2(\epsilon^{(2)}+F)$, where $\eta=\eta_c$; all expressions are similar to the non-driven case.
\textcolor{black}{All previous results hold for systems arbitrarily far from equilibrium. However, some insights can be obtained close to equilibrium. For instance, by introducing the ``forces", $F_\epsilon=(\epsilon^{(1)}-\epsilon^{(2)})/2,F_\beta=(\beta_1-\beta_2)/2$, where we assume $F\ll 1,\,F_\epsilon\ll 1,\,F_\beta\ll 1$ for the two-state case, $\langle{\cal P}\rangle$, $\langle {\dot \sigma}\rangle$ and ${\rm var }({\cal P})$ become $\langle{\cal P}\rangle=(F_\epsilon-F)(\beta F_e+F_{\beta} \epsilon-\beta F){\rm sech}(\beta \epsilon/2)$, $\langle {\dot \sigma}\rangle=(\beta F_e+F_{\beta} \epsilon-\beta F){\rm sech}(\beta \epsilon/2)$ and ${\rm var }({\cal P})=2(F_\epsilon-F)^{2}(\beta \epsilon/2)$, respectively, where  $\epsilon=(\epsilon^{(1)}+\epsilon^{(2)})/2$ and $\beta=(\beta_1+\beta_2)/2$. The linear analysis also shows that the roots of $\langle{\cal P}\rangle$, $F=F_\epsilon$ and $\beta F_\epsilon+F_\beta \epsilon=\beta F$, coincide with the minima of power variance and dissipation, respectively. Although the case of different couplings between thermal baths ($\Gamma_1\neq \Gamma_2$) leads to different expressions for the aforementioned quantities, the roots of power as well as their relation to the minima of ${\langle {\dot \sigma}\rangle}$ and ${\rm var}(\cal P)$ remains the same.}
  
\begin{figure}[t!]
    \centering
    \includegraphics[width=\linewidth]{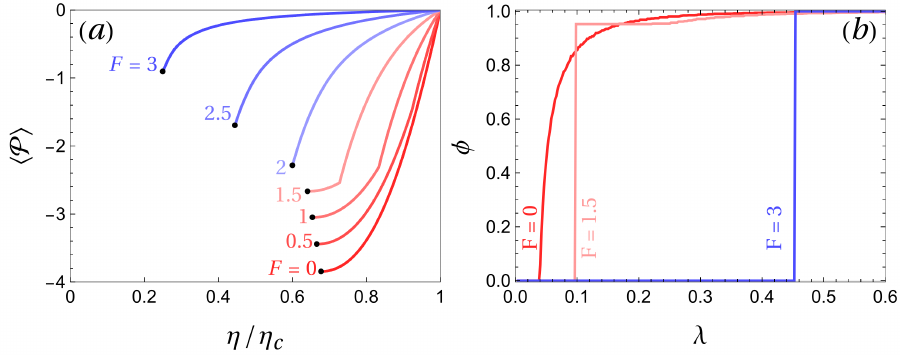}
     \caption{{\bf (a)} Pairwise  $\langle \mathcal{P}\rangle$--$\eta/\eta_c$ 
     Pareto fronts for different non-conservative driving $F$'s (colored numbers). Optimizations have been carried out with respect to individual energies   $\epsilon^{(\nu)}$'s and the temperatures
     $\beta_\nu$'s. Parameter ranges used in the optimization are $\epsilon^{(1)},\,\epsilon^{(2)} \in[-7,0]$, $\beta_1\in [0,2], \,\beta_2 \in [0,\beta_1]$. {\bf (b)} Phase transitions in the order parameter $\phi=\langle {\cal P}\rangle/\langle {\cal P}\rangle_{\rm max}$ of the trade-offs in panel (a), for selected values of $F$.}
    \label{power_efficiency_Twostate_combined}
\end{figure}

All pairwise Pareto fronts exhibit behavior akin to the non-driven case, except when all parameters $\{\epsilon^{(1)},\epsilon^{(2)},\beta_1,\beta_2\}$ are varied and $F$ is fixed and increased, as shown in Fig.~\ref{power_efficiency_Twostate_combined}(a). 
In these cases the fronts acquire a (locally) concave shape, leading to phase transitions in the optimal protocol as a function of $\lambda$. Fig.~\ref{power_efficiency_Twostate_combined}(b) shows this feature for the driven two-state system for different choices of $F$ and order parameter $\phi=\langle {\cal P}\rangle/\langle {\cal P}\rangle_{\rm max}$; for small driving force (darker red lines), Pareto fronts remain convex and the system parameters can be smoothly tuned from the onset of the heat engine regime at $\langle \mathcal{P}\rangle = 0$ and $\phi = 0$ to the state with maximal power where $\phi = 1$, exploring all optimal solutions lying on the front in between these extrema. 
Increasing the driving leads to Pareto fronts exhibiting a locally concave region, preventing a smooth transition from $\phi =0$ to $\phi >0$. For a critical value $\lambda_c$, the optimal protocol jumps from $\phi=0$ to a finite value, after which it smoothly transitions once again to $\phi=1$. Likewise, for even higher $F$ (blue lines in Fig.~\ref{power_efficiency_Twostate_combined}), the Pareto front becomes fully concave and the only optimal designs are the $\phi=0$ state where no power is produced, and the maximal-power state at $\phi=1$; solutions lying in between can only be explored through hysteresis.

\textit{Three-state systems—}  
\textcolor{black}{The (non-interacting) three-state system is similar to
the two-state case, but the single unit can be in either a low ($i=1$), intermediate ($i=2$) or high-energy state ($i=3$), with corresponding energies  $\epsilon_1^{(\nu)}$, $\epsilon_2^{(\nu)}$ and $\epsilon_3^{(\nu)}$, respectively. Expressions for affinities are shown in Appendix~\ref{rates}. This three-state case exhibits} a richer behavior, since the roots of the power output do not necessarily coincide with minima of ${\langle {\dot \sigma}\rangle}$ or ${\rm var}(\cal P)$. 
Despite the existence of six different affinities and fluxes, only  four (affinities) and two (fluxes) are independent. To illustrate this, we first consider non-driven systems.

In this case, expressions for $\langle {\cal P} \rangle$ and $\langle {\dot \sigma}\rangle$ are given by
\begin{equation}
    \label{pow3states}
    \begin{split}
        \langle \mathcal{P} \rangle &= (\Delta \epsilon_{12}^{(1)}-\Delta \epsilon_{12}^{(2)})\,(J_{12}^{(1)} + J_{13}^{(1)}) \\
        &+ (\Delta \epsilon_{23}^{(1)}-\Delta \epsilon_{23}^{(2)})\,(J_{23}^{(1)} + J_{13}^{(1)})
    \end{split}
\end{equation}
and
\begin{equation}
    \label{meanvelocity}
    \begin{split}
        \langle \dot{\sigma} \rangle &= (\beta_1\,\Delta \epsilon_{12}^{(1)}-\,\beta_2\,\Delta \epsilon_{12}^{(2)})\,(J_{12}^{(1)} + J_{13}^{(1)}) \\ 
        &+(\beta_1\,\Delta \epsilon_{23}^{(1)}-\,\beta_2\,\Delta \epsilon_{23}^{(2)})\,(J_{23}^{(1)} + J_{13}^{(1)}),
    \end{split}
\end{equation}

respectively. Although it can be directly obtained from Eqs.~\eqref{var2} or~\eqref{var3}, an expression for the power fluctuations is more cumbersome and will not be explicitly shown here.

\begin{figure}[t!]
    \centering
    \includegraphics[scale=0.25]{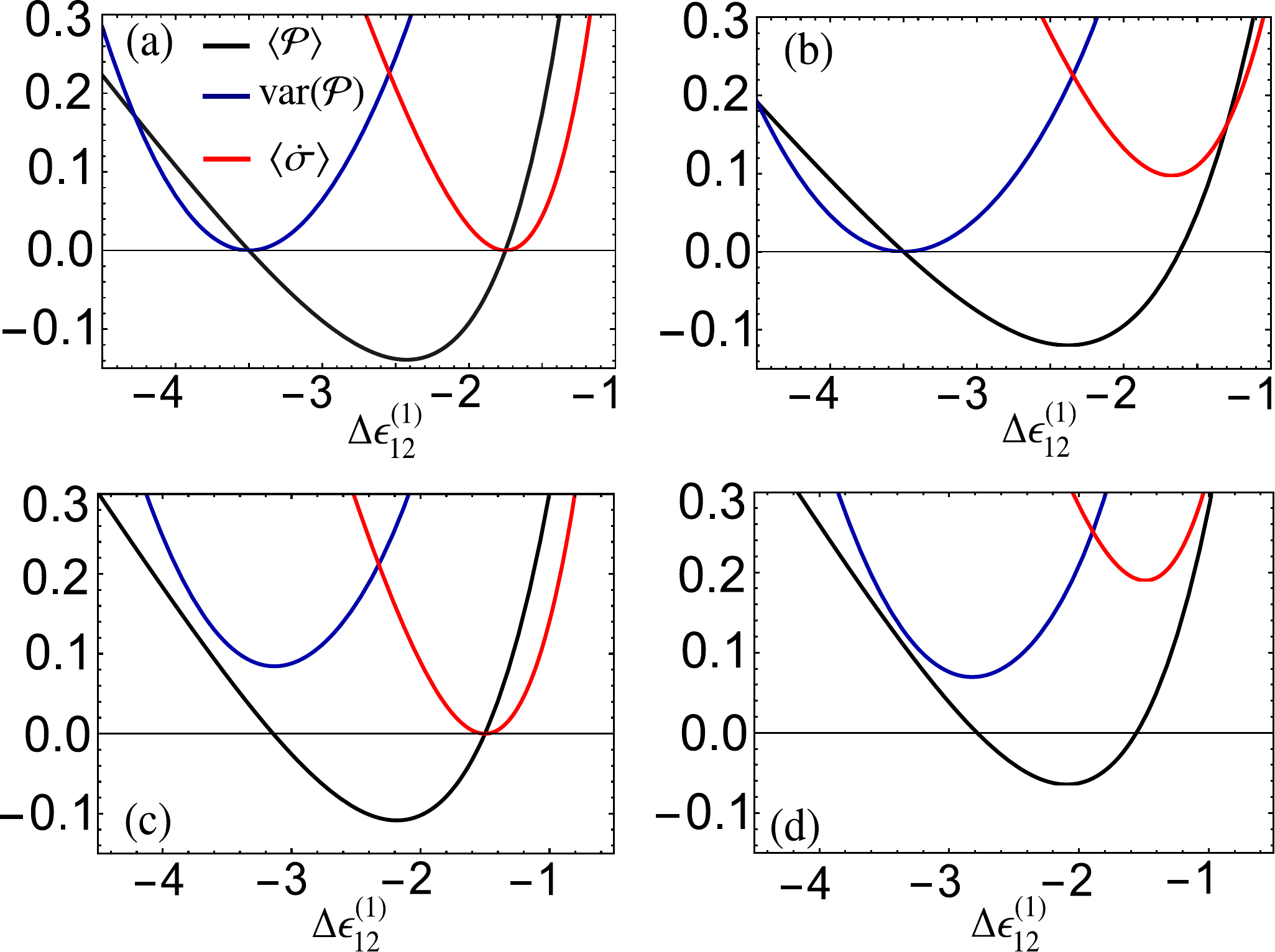}
     \caption{For the three-state system, the
     depiction of $\langle {\cal P}\rangle,\langle {\dot \sigma}\rangle$ and $\textrm{var}\,(\mathcal{P})$  for different set of individual energies. The energies were chosen as follows: $\Delta \epsilon_{ij}^{(\nu)}= (j-i)\,\epsilon^{(\nu)}$, where $\epsilon^{(2)} = -3.5$ (a); $\Delta \epsilon_{12}^{(2)} = \Delta \epsilon_{23}^{(2)} = \Delta \epsilon_{23}^{(1)} = -3.5$  (b); $\Delta \epsilon_{23}^{(1)} = (\beta_2/\beta_1) \Delta \epsilon_{23}^{(2)} = -3$ and $\Delta \epsilon_{12}^{(2)} = -3$ in (c) and $\Delta \epsilon_{12}^{(2)} = -3$, $\Delta \epsilon_{23}^{(1)} = -7$ and $\Delta \epsilon_{23}^{(2)} = -5$ (d).  Parameters: $\beta_1 = 2$ and $\beta_2 = 1$.}
    \label{fig2}
\end{figure}

Unlike the two-state case, there are now four different roots of $\langle {\cal P}\rangle=0$, i.e.,
\begin{align}
\Delta \epsilon_{ij}^{(1)} &= \Delta \epsilon_{ij}^{(2)}, \tag{i}\nonumber\\
(\Delta \epsilon_{12}^{(1)}-\Delta \epsilon_{12}^{(2)})\,(J_{12}^{(1)} + J_{13}^{(1)})&=(\Delta \epsilon_{23}^{(2)}-\Delta \epsilon_{23}^{(1)})\,(J_{23}^{(1)} + J_{13}^{(1)}),  \tag{ii}\nonumber\\
J_{13}^{(1)} &= -J_{12}^{(1)} = -J_{23}^{(1)},  \tag{iii}\nonumber\\
\beta_1\,\Delta \epsilon_{ij}^{(1)} &= \beta_2\,\Delta \epsilon_{ij}^{(2)}\,. \tag{iv}\nonumber
\end{align}
While conditions $\rm (i)$ and $\rm (iv)$ (for all states $i$ and $j$) imply that  $\langle {\dot \sigma}\rangle=0$  and ${\rm var}({\cal P})=0$, respectively (see Fig.~\ref{fig2}(b,c)), the existence of more than two affinities implies that conditions $\rm (ii)$ and $\rm (iii)$ are not necessarily related to the above minima, (Fig.~\ref{fig2}(d)).
%Second, although the condition
%$\Tilde{W}_{ij}\cdot\Tilde{W}_{jk}\cdot\,\cdot\,\cdot\Tilde{W}_{Ni} %=\Tilde{W}_{iN}\cdot\Tilde{W}_{kj}\cdot\,\cdot\,\cdot\Tilde{W}_{ji}$ %($\tilde{W}_{ij}=\sum_\nu W^{(\nu)}_{ij}$)  holds for,
%the existence of more than one affinities
%can lead to different conditions in which $\langle {\cal P}\rangle=0$
%but  not the vanishing of entropy production and/or fluctuations as well as  %their minima.   
However, these minima can still be relatively close to the power zeroes, indicating that the present framework may represent an approximate characterization of the heat engine regime. A special case in which all above conditions are met, regardless of the number of energy levels, is for systems with transitions between neighboring states differing with a fixed amount of energy \(\pm\epsilon^{(\nu)}\) (Fig.~\ref{fig2}$(a)$). In this case, the system power reduces to a simpler form, akin to the two-state system, i.e.,
\begin{equation}
\langle \mathcal{P}\rangle=(\epsilon^{(1)} - \epsilon^{(2)}) \sum_{i<j} (j-i) J_{ij}^{(1)}.
\end{equation}

Before proceeding to collective systems, we address some results in the presence of non-conservative drivings $F^{(\nu)}_{ij}\neq 0$. Among the different possibilities, we also follow the ideas of Refs.~\cite{filho2023powerful,splitting2024} in which the driving stems from a biased force favoring the following sequence of transitions between the states: $1\rightarrow 2\rightarrow 3\rightarrow 1$ and $1\rightarrow 3\rightarrow 2\rightarrow 1$ according to whether the transition \( j \rightarrow i \) is associated with the cold or hot thermal bath, respectively,
as sketched in Fig.~\ref{scheme}. Mathematically, $F^{(\nu)}_{ij}$ is given by $-F$ provided $\mod(i-j,3)=1$ and $\mod(i-j,3) = 2$ in the cases where the transition is associated with the cold and hot thermal bath, respectively, and $F$ otherwise.

Unlike the two-state case, the power zeros do not necessarily coincide with the minima of entropy production and fluctuations when $F^{(\nu)}_{ij} \neq 0$, \textcolor{black}{as is the case for exact calculations in the linear analysis (not shown)}. This follows from a simple argument: the presence of non-conservative forces implies that the condition $\prod_{\gamma} W_{ij}^{(1)} \cdot \prod_{\gamma} W_{ij}^{(2)} = 1$, which is necessary for $J_{ij}^{(1)} = -J_{ij}^{(2)}$, no longer holds. Indeed, for closed paths in the same direction, this product becomes $e^{\pm \frac{3}{2} F(\beta_1 - \beta_2)} \neq 1$, and for opposite directions, it becomes $e^{\pm \frac{3}{2} F(\beta_1 + \beta_2)} \neq 1$. This shows that the condition is fulfilled only when $F^{(\nu)}_{ij} = 0$. Nevertheless, the results concerning Pareto fronts remain qualitatively similar to those in the two-state case and will hence not be shown here.

\subsection{Interacting systems}\label{interacting}

In this section, we extend our analysis to heat engines operating collectively. Systems presenting collective behavior are of central importance in several areas of equilibrium and nonequilibrium statistical physics and have attracted recent attention in the realm of stochastic \cite{filho2023powerful,herpich,herpich2,splitting2024,hawthorne2023nonequilibrium,mamede2023} and quantum-thermodynamics \cite{PhysRevA.107.L040202} as reliable approaches for boosting the performance of heat engines. Unlike the previous examples, here we consider a simple model \cite{filho2023powerful,splitting2024} in which both interaction and individual energies assume the same values when in contact with both thermal reservoirs. The system is composed of $N$ interacting units, where each unit \( i \) is represented by a spin variable \( s_j \) assuming the values \( s_i \in \{-1, 0, 1\} \); a given microscopic state is then defined by the configuration of individual spins, i.e., \( s \equiv (s_1, \ldots, s_i, \ldots, s_N) \). 
The energy of the system takes the simple Ising-like expression
\begin{equation}
E(s) = \frac{\epsilon}{2k} \sum_{(i, j)} s_i s_j + \Delta \sum_{i=1}^{N} s_i^2\,,
\label{ising}
\end{equation}  
where $k$ quantifies the number of nearest neighbor spins and the first and second terms on the right hand side account for the interaction between two nearest neighbor units (with strength \( \epsilon \)) and the individual energy, respectively. The latter assumes the values \(0\) or \( \Delta \) provided $s_i=0$ or $s_i\neq 0$, respectively. \textcolor{black}{ Each configuration change, say from  \( s  \) to \( {\tilde s} \equiv (s_1, \ldots, {\tilde s}_i, \ldots, s_N) \), is characterized by a spin flip from $s_i \in \{-1, 0, 1\}$ to ${\tilde s}_i \in \{-1, 0, 1\}$, where  $s_i\neq {\tilde s}_i$.
 }

 \begin{figure*}[htp]
     \centering
     \includegraphics[width=0.9\linewidth]{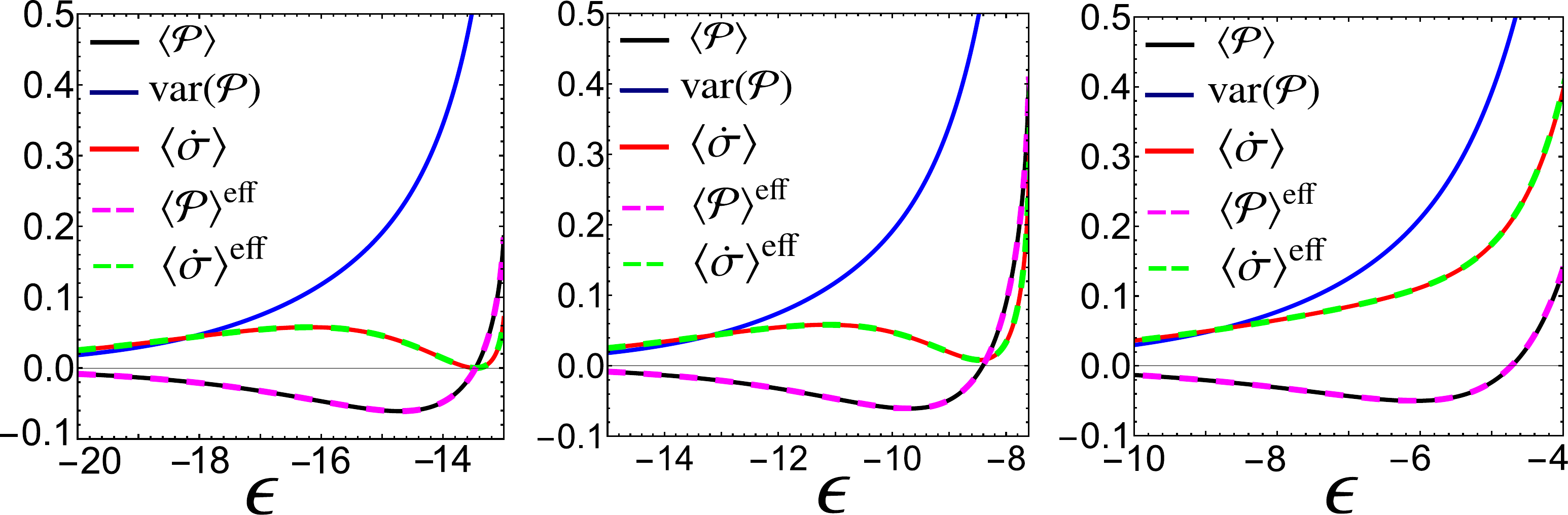}
     \caption{Depiction of $\langle {\cal P}\rangle,\langle {\dot \sigma}\rangle$ and $\textrm{var}\,(\mathcal{P})$ versus
     $\epsilon$ for different $\Delta$'s ($\Delta=10,5$ and $1$ from the
     left to right panels).      
     The superscript `eff' refers to the ``effective" description, as discussed in Section \ref{pfd}.
     \textcolor{black}{The horizontal black line at $\langle\mathcal{P}\rangle=0$ is shown for clarity, clearly showing the roots.}  
     Parameters: $F = 1.1$, $\beta_1 = 2$, and $\beta_2 = 1$.}
     \label{collective}
 \end{figure*}

As in previous examples, each unit is also simultaneously coupled with two thermal baths and subjected to external driving \( F^{(\nu)}_{ij} \) favoring certain
spin transitions. This driving acts on the local spin change \( s_j \rightarrow \tilde{s}_j \) in a similar fashion as the three-state model, reading $-F$ provided $\mod(\tilde{s}_j-s_j,3)=1$ and $\mod(\tilde{s}_j-s_j,3)=2$ in the case the local spin change is associated  with the cold and hot thermal baths, respectively, and $F$ otherwise, as shown in Fig.~\ref{scheme}.
%to a clockwise or a counterclockwise bias depending
%on the thermal bath.
%Clockwise transitions are defined as those for which an individual state changes according to the sequence \( -1 \rightarrow 0 \rightarrow +1 \rightarrow -1 \), while counterclockwise transitions correspond to \( +1 \rightarrow 0 \rightarrow -1 \rightarrow +1 \). 
%according to the clockwise (\(-1 \to 0 \to +1\to -1\)) or counterclockwise (\(+1 %\to 0 \to -1 \to 1\)) direction, depending on the thermal bath the system
%is coupled. The former and latter  set of transitions are favored when the system is coupled
%with the cold and hot thermal reservoirs, respectively.
Since the model is not exactly solvable for regular lattice topologies, we consider its all-to-all description $k \rightarrow N$. The dynamics are fully described via the total number $N_i$ of spins in each state $i \in \{0,\pm\}$, with $N_++N_-+N_0=N$. The system energy then becomes
\begin{equation}
    \label{mfeq}
    \begin{split}
        E(s)&\rightarrow\frac{\mathcal{\epsilon}}{2N}\left\{ N_+(N_+-1)+N_-(N_--1) -2 N_+N_- \right\} \\
        &+\Delta\left(N_++N_-\right).
    \end{split}
\end{equation}
For $N\rightarrow \infty$, the dynamics is characterized by 
the density of states, \( p_{i} = \langle N_{i}/N \rangle \), {with} \( i \in \{0,\pm\} \), the time evolution of which is governed by a master equation akin to Eq.~\eqref{me}, i.e.,
\begin{equation}
    \dot{p}_i(t) = \sum_{\nu=1}^2 \sum_{j \neq i}  W^{(\nu)}_{ij} p_{j}(t) - W^{(\nu)}_{ji} p_i(t)\,.
\end{equation}

\noindent \textcolor{black}{The transition rates $W_{ij}^{(\nu)}$ are listed in Appendix \ref{rates}}. As for the three-state, the non-conservative driving  implies that $\prod_{\gamma} W_{lm}^{(1)} \cdot \prod_{\gamma} W_{lm}^{(2)} = e^{\pm \frac{3}{2} F(\beta_1 - \beta_2)}$  and $\prod_{-\gamma} W_{lm}^{(1)} \cdot \prod_{-\gamma} W_{lm}^{(2)} = e^{\pm \frac{3}{2} F(\beta_1 + \beta_2)} $ for paths in the same and  opposite direction, respectively. As such, the roots of the power will not necessarily coincide
with the minimum of power fluctuation and dissipation. 
However, for large $ -\epsilon$ and $\Delta$ and for $F\ll 1$, the system behavior approximates the two-state dynamics because the dominated dynamics occur between 
transitions $-$ and $0$. In such case, $\langle \mathcal{P}\rangle $, ${\rm var}({\cal P})$ and $\langle \sigma\rangle$ can be evaluated from the assumption $J_{- \leftrightarrow 0}^{(1)} = -J_{- \leftrightarrow 0}^{(2)}$, since  $-\leftrightarrow 0$ are the dominant transitions \textcolor{black}{even if the $\Gamma_i$'s are different, as discussed previously for the two-state case}. Such a description is fully equivalent to the phenomenological description developed in Refs.~\cite{filho2023powerful,splitting2024}, whose main expressions for the different thermodynamic quantities are approximately given by
\begin{equation}
\langle \mathcal{P}\rangle \approx \frac{2 F \left(e^{\frac{1}{2} \beta_1 (\Delta + \epsilon + F)} - e^{\frac{1}{2} \beta_2 (\Delta + \epsilon - F)}\right)}{e^{\frac{1}{2} ((\beta_1 + \beta_2) (\Delta + \epsilon) + F (\beta_1 - \beta_2))} + 1},
\label{prodef}
\end{equation}

\begin{equation}
{\rm var}({\cal P})\approx 4F^2 \left[e^{\frac{\beta_1}{2} (\epsilon +\Delta+ F)} + e^{\frac{\beta_2}{2} (\epsilon+\Delta - F)}\right], 
\end{equation}
and
\begin{equation}
\langle \sigma\rangle \approx e^{\frac{1}{2} \beta_2 (\Delta - F + \epsilon)} [\beta_2(\Delta+\epsilon-F)-\beta_1(\Delta+\epsilon+F)].
\end{equation}
It can be seen that, while the approximate power $\langle \mathcal{P}\rangle$ vanishes as either $F=0$ or $\beta_1(\Delta + \epsilon + F)=\beta_2 (\Delta + \epsilon - F)$, the former and latter roots are solutions of ${\rm var}({\cal P})=0$ and $\langle \sigma\rangle=0 $, respectively. Fig.~\ref{collective} depicts such findings for different $\Delta$ as $\epsilon$ is varied (the remaining parameters,  $F$ and $\beta_\nu$, are fixed). In this case, while the engine regime is approximately constrained between $\epsilon \rightarrow -\infty$ and $\beta_1(\Delta + \epsilon + F)=\beta_2 (\Delta + \epsilon - F)$ for large and intermediate $\Delta$, it deviates from the above solutions as $\Delta$ is small.

\begin{figure}[htp]
    \centering
    \includegraphics[width=\linewidth]{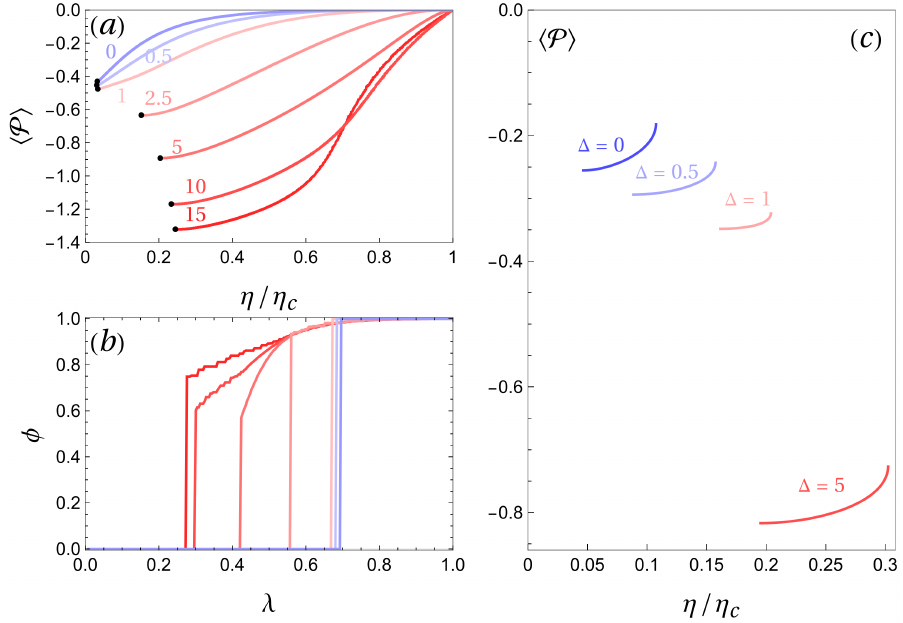}
    \caption{{\bf (a)} Power-efficiency Pareto trade-off for the interacting system with fixed $F=1$ and varying $\Delta$ (numbers). Red and blue lines indicate trade-offs for $\Delta \geq F$ and $\Delta < F$, respectively. Parameters: {\bf (b)} Phase transitions in the optimal parameters for the power-efficiency trade-off. For $\Delta \leq F$, $\langle\mathcal{P}\rangle$ switches between zero and $\langle\mathcal{P}\rangle_{\rm max}$ (black dots in (a)) at a given critical $\lambda_c$, whereas   $\langle\mathcal{P}\rangle$ switches between 0 and another value that is not $\langle\mathcal{P}\rangle_{\rm max}$ for $\Delta > F$. In this case, it smoothly transitions to it for $\lambda>\lambda_c$ due to the local convexity of the front.
    {\bf (c)} Similar to panel $(a)$, but for $F=1, \,\beta_1=5,\,\beta_2=1$ and fixed but tuneable $\Delta$. Unlike $(a)$, all power-efficiency Pareto trade-offs
    are convex.}
    \label{fig:P_efficiency_collective}
\end{figure}

Lastly, we discuss some similar as well as different aspects about the Pareto fronts found in the collective system. While $\langle {\cal P}\rangle-\eta/\eta_c$ Pareto fronts are convex when the majority of the system parameters $\{\Delta,\beta_1,\beta_2,F\}$ are held fixed  [Fig.~\ref{fig:P_efficiency_collective}c],
they exhibit (locally) concave shapes when the majority of parameters are varied (Fig.~\ref{fig:P_efficiency_collective}a),
consistent with smooth (not shown) and discontinuous behaviors of the order parameter $\phi$, respectively. However, there is a remarkable difference with respect to the two-state system, as depicted in Fig.~\ref{fig:P_efficiency_collective}b. There is no change of concavity in the Pareto fronts associated with $\Delta \leq F$, meaning that in the optimal engine design $\langle\mathcal{P}\rangle$ switches between zero ($\phi=0$) and $\langle\mathcal{P}\rangle_{\rm max}$ ($\phi=1$) at a given critical $\lambda_c$. Conversely, the $\langle {\cal P}\rangle-\eta/\eta_c$ Pareto fronts
present an inflection point for $\Delta > F$, consistent with $\phi$ jumping at $\lambda=\lambda_c$ to a value smaller than $1$. In such cases, $\langle\mathcal{P}\rangle$ switches between 0 and  $\langle\mathcal{P}\rangle<\langle\mathcal{P}\rangle_{\rm max}$ for $\lambda=\lambda_c$, smoothly approaching $\langle\mathcal{P}\rangle_{\rm max}$ as $\lambda$ increases.

\section{Conclusions}\label{conclusions}
We introduced an alternative approach for the characterization of nonequilibrium heat engines of systems placed in contact with two thermal reservoirs, based on the relation between the roots of the power output and the minima of dissipation and power fluctuations. \textcolor{black}{This relation is exact for generic two-state systems and can be approximately verified in systems with an arbitrary number of levels, hence providing a good description of the behavior of general systems exhibiting a dominant dynamics  which can be approximately treated as  a two-state system}. Such a description has been shown to be useful for locating the engine regime, in order to consider different strategies for the optimization of heat engines. 
We advanced beyond the individual maximizations of power and efficiency by introducing distinct types of Pareto fronts. They constitute a remarkable approach for studying the class of heat engines treated in this work. Results have shown that the existence of asymmetries in the dynamics (non-conservative driving forces in our case) not only influences the strength of  optimization but also its shape. While Pareto fronts are convex for small drivings strengths in which the power smoothly changes from zero (absence of a heat engine regime) to its maximum value, the fronts' concave shapes for higher driving values reveal an interesting feature, akin to  discontinuous phase transitions, in which the power  abruptly switches from its root value to its maximum value. 
\textcolor{black}{It is insightful mentioning the relationship between our findings
and the TURs. In particular, the minimum of entropy production is featured
by all net heat currents vanishing  ($\langle\dot{Q}_v\rangle \to 0$) and 
hence the condition $\text{var}(J)\langle\dot{\sigma}\rangle \ge 2\langle J \rangle^2$ is identically satisfied. Conversely, the minimum of power fluctuations corresponds to a nonequilibrium state where the system  operation resembles an ideal heat conductor in which $\langle\dot{Q}_1\rangle = -\langle\dot{Q}_2\rangle$. Since this is a nonequilibrium state with positive total entropy production, the TUR for the underlying heat currents holds as a general inequality. Further, the TUR provides a fundamental bounding curve in the space of fluctuations vs. dissipation (and hence also vs. power or efficiency), and when one sets up a multi‐objective optimization, the resulting Pareto front necessarily either touches or follows this TUR‐derived bound in those regions where fluctuations are a limiting factor~\cite{berx2024universal}.}
Finally, it is worth highlighting some potential directions for future research. In particular, it would be interesting to characterize and optimize heat engines in other classes of systems—especially \textcolor{black}{those described by periodically driving systems \cite{PhysRevB.99.224306,PhysRevLett.114.183602,PhysRevE.93.042112,harunari2020maximal} and also those} involving non-simultaneous contact with thermal reservoirs—and to explore the resulting Pareto front structures in such scenarios.

\section{Acknowledgments}
G.A.L.F and C.E.F. acknowledge the financial support from FAPESP under grants 2022/15453-0, 2022/16192-5 and 2024/03763-0. The financial support from CNPq is also acknowledged. J.B. is supported by the Novo Nordisk Foundation with grant No. NNF18SA0035142.

\bibliography{refs}

%\newpage
%\clearpage

%\appendix            

%\onecolumngrid
%\begin{center}
%\textbf{\large Supplemental Material: Power, %dissipation and fluctuation relations in collective %heat engines}
%\label{sup}
%\end{center}

%\begin{center}
% Gustavo A. L. Forão, Jonas Berx and C. E. Fiore
%\end{center}

\appendix
\section{Main expressions for covariances}\label{covv}
In order to compute $J_{ij}^{(\nu)}$ and $C^{({\nu},{\nu'})}_{iji'j'}$ from
the transition rates in the steady-state regime, it is convenient
to introduce the incremental time $\Delta t$ in such a way
 that $t_f$ and $M$ are related via the relation
$t_f=M\Delta t$. From this,   $ \langle n^{(\nu)}_{ij}\rangle$
and $\langle n^{(\nu)}_{ij} n^{(\nu')}_{i'j'} \rangle$ read
$\langle n^{(\nu)}_{ij}\rangle=t_f W^{(\nu)}_{ij}p^{\rm st}_{j}$ 
and
% \begin{gather}
% \label{meanME}
% \langle n^{(\nu)}_{ij} n^{(\nu')}_{i'j'} \rangle = W^{(\nu)}_{ij} W^{(\nu')}_{i'j'} \int_0^{t_f} dt \int_0^{t} d\tau \Big( p_{j';t| i; \tau} p_j^{\rm st} + \nonumber \\ 
% + p_{j; t| i'; \tau} p_{j'}^{\rm st} \Big) + \delta_{\nu\nu'}\delta_{ii'} \delta_{jj'} t_f W_{ij} p_j^{\rm st},
% %\CEF{\langle n_{ml} n_{i'j'} \rangle = W_{ml} W_{i'j'} \int_0^{t_f} dt \Big(  \int_t^{t_f} d\tau  p_{l';t| m; \tau} p_l^{\rm st} }+ \nonumber \\ 
% %\CEF{+  \int_0^{t} d\tau p_{l; t| m'; \tau} p_{l'}^{\rm st} \Big) + \delta_{mm'} \delta_{ll'} t_f W_{ml} p_l^{\rm st}.}
% \label{sdevME}
% \end{gather}
\begin{widetext}
\begin{equation}
    \label{meanME}
    \begin{split}
        \langle n^{(\nu)}_{ij} n^{(\nu')}_{i'j'} \rangle = W^{(\nu)}_{ij} W^{(\nu')}_{i'j'} \int_0^{t_f} dt \int_0^{t} d\tau \Big( p_{j';t| i; \tau} p_j^{\rm st} + p_{j; t| i'; \tau} p_{j'}^{\rm st} \Big)+ \delta_{\nu\nu'}\delta_{ii'} \delta_{jj'} t_f W_{ij} p_j^{\rm st},
    \end{split}
\end{equation}
\end{widetext}
respectively. By taking the continuous time limit $\Delta t\rightarrow 0$ for sufficiently long times $t_f\rightarrow \infty$, we obtain  the expression for $\langle \dot{Q}_\nu \rangle$ introduced previously by inserting the first average  
into Eq.~(\ref{av1}). By proceeding in a similar fashion for $\langle n^{(\nu)}_{ij} n^{(\nu')}_{i'j'} \rangle $, we find that the asymptotic expression for $C^{(\nu,\nu')}_{iji'j'}/t^2_f$ becomes
\begin{gather}
    \label{JtildeLTL}
    \frac{C^{(\nu,\nu')}_{iji'j'}}{t_f^2} \rightarrow W^{(\nu)}_{ij} W^{(\nu')}_{i'j'} \bigg( p^{\rm st}_{j'} \int_0^{+\infty} dt \left( p_{j; t| i'; 0} - p^{\rm st}_j \right) + \nonumber\\
    + ~p^{\rm st}_{j} \int_0^{+\infty} dt \left( p_{j'; t| i; 0} - p^{\rm st}_{j'} \right) \bigg) + \delta_{\nu\nu'}\delta_{ii'} \delta_{jj'} W^{(\nu)}_{ij} p^{\rm st}_j,
    \label{CtildeLTL}
\end{gather}
where we have taken into account that the above conditional probabilities only depend on time differences in the steady state.
%. Note that $C^{({\nu},{\nu'})}_{iji'j'}$ depends
%on the conditional probabilities of type $p_{j; t| i'; 0}$  of the system be in the state
%$i$ at time $t$ given it was in the state $j'$ at $t=0$.
Since we are consider Markovian systems, $ p_{i; t| j'; 0}$ can be evaluated by expanding all probabilities in terms of the eigenvalues and eigenvectors of the total transition matrix, i.e.,
\begin{equation}
p_{j; t| i; 0} = p_j^{\rm st} + \sum_{\ell=2}^N v_j^{(\ell)} a_\ell^{(i)} e^{{\bar\lambda}_\ell t},
\label{cond}
\end{equation}
where $N$ is the total number states in the system, $v_i^{(\ell)}$ is the $i$-th component of the $\ell$-th eigenvector, and ${\bar \lambda}_\ell$ its associated eigenvalue, enumerated in descending order ($\lambda_1 = 0 > \lambda_2 > \dots > \lambda_N$).  By inserting Eq.~(\ref{cond}) into Eq.~(\ref{CtildeLTL}), ${\rm var}(\cal P)$ is finally evaluated.
%\[\\
%p_{i; t| j; 0} = p_i^{\rm st} + \sum_{\ell=2}^N v_i^{(\ell)} a_\ell^{(j)} %e^{{\bar\lambda}_\ell t},\]

In the second approach, the power variance is evaluated via the characteristic function for the power combined with the large deviation method \cite{touchette2009large, kumar2011thermodynamics}. Although no simple expression is obtained in general, it provides a rather simple and straightforward tool for evaluating the power variance, which is equivalent to Eq.~(\ref{var2}).
Let \( P(i, \mathcal{P}, t) \) denote the probability of the system being in state \( i \) at time \( t \) with power \( \mathcal{P} \). The time evolution of \( P(i, \mathcal{P}, t) \) is governed by the master equation 
\begin{widetext}
    \begin{equation}
        \frac{\partial}{\partial t} P(i, \mathcal{P}, t) = \sum_{\nu=1}^2 \sum_{j\neq i} \left\{ W_{ij}^{(\nu)} P(j, \mathcal{P} - \Delta \mathcal{P}, t) - W_{ji}^{(\nu)} P(i, \mathcal{P}, t) \right\},    
    \end{equation}
\end{widetext}

where  \( \Delta \mathcal{P} \) denotes the amount
of power due to the transition between states \( j \) and \( i \), 
The characteristic function of the power is defined as \( \rho_p(i, \alpha, t) = \int_{-\infty}^{\infty} d\mathcal{P} \, e^{-\alpha \mathcal{P}} P(i, \mathcal{P}, t) \), whose time evolution is described by  
\begin{widetext}
    \begin{equation}
        \frac{\partial}{\partial t} \rho_p(i, \alpha, t) = \sum_{\nu=1}^2 \sum_{j} \left\{ W_{ij}^{(\nu)} \rho_p(j, \alpha, t) e^{-\alpha \Delta \mathcal{P}} - W_{ji}^{(\nu)} \rho_p(i, \alpha, t) \right\},
    \end{equation}
\end{widetext}
which can be rewritten in a more compact form,
\begin{equation}
    \frac{\partial \rho_p(i, \alpha, t)}{\partial t} = M_p(\alpha) \rho_p(i, \alpha, t)\,, 
\end{equation}
with \( M_p(\alpha) \) the tilted stochastic generator. Using a large-deviation principle~\cite{touchette2009large, kumar2011thermodynamics}, the scaled cumulant generating function is determined by the largest eigenvalue \( \lambda_p(\alpha) \) of \( M_p(\alpha) \). From \( \lambda_p(\alpha) \), the variance is then given by
\begin{equation}
    \textrm{var}({\cal P}) = \left. \frac{\partial^2 \lambda_p(\alpha)}{\partial \alpha^2} \right|_{\alpha=0}.
\end{equation}
Since the expression for \( \lambda_p(\alpha) \) is generally cumbersome, we instead use an alternative approach. In this method, we expand the characteristic function as $a_0 + \sum_{n=1}^N a_n \lambda^n_p(\alpha) = 0$. As shown in Ref.~\cite{wachtel2015fluctuating}, this allows us to write $\textrm{var}({\cal P})$ as  
\begin{equation}
 \textrm{var}({\cal P})= \frac{1}{a_1^3} \left[ a_1^2 \frac{\partial^2 a_0}{\partial \alpha^2} - 2a_1 \frac{\partial a_0}{\partial \alpha} \frac{\partial a_1}{\partial \alpha} + 2a_2 \left( \frac{\partial a_0}{\partial \alpha} \right)^2 \right]_{\alpha=0}.
 \label{var3}
\end{equation}

\section{Proof of the reversibility condition}
\label{proof}

Let us consider  a generic system with arbitrary number of states in which the following equality (Kolmogorov's criterion) holds.
\begin{equation}
    \Tilde{W}_{ij}\cdot\Tilde{W}_{jk}\cdot\,\cdot\,\cdot\Tilde{W}_{Ni} =\Tilde{W}_{iN}\cdot\Tilde{W}_{kj}\cdot\,\cdot\,\cdot\Tilde{W}_{ji}\,.
\end{equation}

Dividing both sides by $\Tilde{W}_{iN}\cdot\Tilde{W}_{kj}\cdot\,\cdot\,\cdot\Tilde{W}_{ji}$, and using the fact that $\Tilde{W}_{lm} = W_{lm}^{(1)} + W_{lm}^{(2)} $ and $W_{lm}^{(\nu)} = 1/W_{ml}^{(\nu)}$ (a consequence of $d_{lm}^{(\nu)} = -d_{ml}^{(\nu)}$), one arrives at the following relation
\begin{equation}
    \frac{W_{ij}^{(1)} + W_{ij}^{(2)}}{\frac{1}{W_{ij}^{(1)}} +\frac{1}{W_{ij}^{(2)}}}\,\cdot\,\frac{W_{jk}^{(1)} + W_{jk}^{(2)}}{\frac{1}{W_{jk}^{(1)}} +\frac{1}{W_{jk}^{(2)}}}\,\cdot \cdot \cdot \,\frac{W_{Ni}^{(1)} + W_{Ni}^{(2)}}{\frac{1}{W_{Ni}^{(1)}} +\frac{1}{W_{Ni}^{(2)}}} = 1\,.
\end{equation}

The above relation can be rearranged as
\begin{equation}
W_{ij}^{(1)}\,W_{ij}^{(2)}\,\cdot\,W_{jk}^{(1)}\,W_{jk}^{(2)}\,\cdot \cdot \cdot\,W_{Ni}^{(1)}\,W_{Ni}^{(2)} = 1\,,
\end{equation}
or equivalently as
\begin{equation}
\prod_{\gamma}W_{lm}^{(1)}\,\cdot\,\prod_{\gamma}W_{lm}^{(2)} = 1,
\end{equation}
as shown in the main text. For systems where $d_{ij}^{(\nu)} = -d_{ji}^{(\nu)}$, this relation is sufficient to ensure the validity of $J_{ij}^{(1)} = -J_{ij}^{(2)}$.

{\color{black}\section{The NSGA-II algorithm}
\label{app:NSGA}
To numerically compute the Pareto fronts in this manuscript, we use the Non-dominated Sorting Genetic Algorithm NSGA-II~\cite{pareto}, implemented in Matlab. It makes use of three core components: elitism, fast non-dominated sorting and a crowding-distance mechanism. A solution is said to dominate another if at performs at least as good in all objectives, and \emph{strictly} better in at least one objective. An initial parent population of $N$ solutions is randomly created, evaluated on all objectives and then sorted into fronts $F$ according to their dominance: $F_0$ contains all non-dominated solutions, $F_1$ all solutions that are only dominated by points on $F_0$ and so on. Within each front the algorithm then computes a crowing measure (Manhattan distance), to estimate local solution density. 

A mating pool is then created by performing a binary tournament selection where two solutions are compared first according to their rank (e.g., $F_0$, $F_1$, etc.) and, if equal rank, according to their crowding distance, where preference is given to larger distances, meaning that solutions will be better spread out across the Pareto fronts. Next, crossover and mutation genetic operators are applied to this mating pool, where the former combines two parent solutions into two offspring solutions by generating values symmetrically around the parents using a probability distribution controlled by a crossover distribution index $\eta_c$; The higher $\eta_c$, the closer the offspring are to the parents. After crossover, a mutation operation perturbs each solution slightly, with the size of the perturbation governed by the mutation distribution index $\eta_m$. This mutation simulates a polynomial probability distribution favoring small changes.

Parents and children are then recombined into a single population and again sorted according to the dominance criterion and crowding distance. Only the $N$ best solutions are retained (elitism) and the procedure restarts. This is iterated until a stopping criterion is met and the final first non-dominated front $F_0$ is returned.}

\section{Transition rates}\label{rates}
{\color{black}
In all cases,  transition rates are generally defined as 
\begin{equation}
    W^{(\nu)}_{ij}=\Gamma\,e^{-\frac{\beta_\nu}{2} d_{ij}^{(\nu)}},
\end{equation}
where $d_{ij}^{(\nu)} = \Delta \epsilon_{ij}^{(\nu)} + F_{ij}^{(\nu)}$.  For the three-state case, 
affinities
$d_{ij}^{(\nu)}$ are given by
\begin{eqnarray}
d_{21}^{(1)}&=&(\epsilon^{(1)}_2- \epsilon^{(1)}_1)- F,\nonumber\\
d_{32}^{(1)}&=&(\epsilon^{(1)}_3- \epsilon^{(1)}_2)- F,\nonumber\\
d_{13}^{(1)}&=&(\epsilon^{(1)}_1- \epsilon^{(1)}_3)- F,
\end{eqnarray}
for $\nu=1$ and
\begin{eqnarray}
d_{21}^{(2)}&=&(\epsilon^{(2)}_2- \epsilon^{(2)}_1)+ F,\nonumber\\
d_{32}^{(2)}&=&(\epsilon^{(2)}_3- \epsilon^{(2)}_2)+ F,\nonumber\\
d_{13}^{(2)}&=&(\epsilon^{(2)}_1- \epsilon^{(2)}_3)+ F,
\end{eqnarray}
for $\nu=2$, respectively. Conversely,
for interacting systems, the transition rates are obtained from Eq.~(\ref{mfeq}) and become probability dependent, with affinities
$d_{ij}^{(\nu)}$ given by
\begin{eqnarray}
d_{-+}^{(1)}&=&2\epsilon(p_- - p_+) - F,\nonumber\\
d_{0+}^{(1)}&=&-\epsilon(p_+ - p_-) - \Delta + F,\nonumber\\
d_{-0}^{(1)}&=&-\epsilon(p_+ - p_-) + \Delta + F,
\end{eqnarray}
for $\nu=1$ and
\begin{eqnarray}
d_{-+}^{(2)}&=&2\epsilon(p_- - p_+) + F\nonumber\\
d_{0+}^{(2)}&=&-\epsilon(p_+ - p_-) - \Delta - F\nonumber\\
d_{-0}^{(2)}&=&-\epsilon(p_+ - p_-) + \Delta - F,
\end{eqnarray}
for $\nu=2$, respectively. The remaining affinities are found by substituting $d_{ij}^{(\nu)}\rightarrow -d_{ji}^{(\nu)}$ in the above expressions.}

\end{document}